\RequirePackage{fix-cm}

\documentclass[smallextended]{svjour3}
      
\usepackage{epsfig}
\usepackage{graphicx} 
\usepackage{dcolumn}  
\usepackage{bm}       
\usepackage{color}
\usepackage{epsfig}
\usepackage{subfigure}
\usepackage{epstopdf}

\def\qsqrt{{\sqrt{2} \kern-1.2em ^4}}
\def\CC{{\rm\kern.24em \vrule width.04em height1.46ex depth-.07ex
\kern-.30em C}}
\def\P{{\rm I\kern-.25em P}}
\def\RR{{\rm
         \vrule width.04em height1.58ex depth-.0ex
         \kern-.04em R}}
\def\id{{\rm 1\kern-.22em l}}
\def\ZZ{{\sf Z\kern-.44em Z}}
\def\NN{{\rm I\kern-.20em N}}

\begin{document}
\title{Topological order in 1D Cluster state protected by symmetry}

\author{W. Son \and  L. Amico \and V. Vedral}

\institute{W. Son \at Department of Physics, Sogang University,Mapo-gu, Shinsu-dong, Seoul 121-742, Korea \\ \email{sonwm@physics.org} \and 
L. Amico \at CNR-MATIS-IMM $\&$ Dipartimento di Fisica e Astronomia Universit\'a di Catania, C/O ed. 10, viale A. Doria 6
95125 Catania, Italy \and V. Vedral \at Center for Quantum Technology, National University of Singapore, 117542 Singapore, Singapore \\  
Department of Physics, National University of Singapore, 2 Science Drive 3, Singapore 117542\\ Department of Physics, National University of Singapore, 2 Science Drive 3, Singapore 117542}



\maketitle

\begin{abstract}
We demonstrate how to construct the $Z_2\times Z_2$ global symmetry which protects the
ground state degeneracy of cluster states for open boundary conditions. Such a degeneracy ultimately arises because the set of stabilizers do not span a complete set of integrals of motion of the cluster state Hamiltonian for open boundary conditions. By applying control phase transformations, our construction makes the stabilizers into the Pauli operators spanning the qubit Hilbert space from which the degeneracy comes.
\end{abstract}

\section{Introduction}
Measurement based quantum computation \cite{Briegel01,Raussendorf03,Nielsen05} provides an important scheme among the diverse implementations proposed so
far \cite{Nielsen00,Adiabatic}. Its computational power is encoded in the cluster states, where entanglement and, more generically, quantum correlations are the core ingredients of the circuit. Topological features are demonstrated to provide a fault tolerance of the state that is important for qubit manipulations to fight against decoherence \cite{Sarma05}. Indeed, the notion of  topological order exploited in this context comes from quantum statistical physics. Namely, it is a specific property of the  ground state (gs) in certain two dimensional systems \cite{boundary,kitaev-leshouches} that acquires a degeneracy as tightly related to the boundary conditions; such degeneracy must be robust under arbitrary local perturbations. On the contrary, two dimensional cluster states does not have such topological order \cite{2d-cluster}. In the present paper we focus on one dimensional (1d) systems.  We shall see that, despite of the limitation of computational power in this case \cite{Nest06,Damian07}, 1d cluster states could constitute an interesting platform for the computation because they enjoy a global order of topological origin.

The notion of topological order in 2d systems cannot be directly applied to 1d as it stands. In fact, linear chain gs can well display a degeneracy which is sensitive to boundary conditions, but local operators do exist in this case that can disturb the gs degeneracy.
Indeed, it was clarified recently that the notion of topological order in 1d is possible only with the assistance of the symmetry protection \cite{Wen-protected}. In this context it was demonstrated how, without symmetry, any quantum phase in 1d (with a finite energy gap) can be adiabatically continued to a fully polarized phase (displaying, of course, no topological order). This allows us to endow a more precise meaning to the notion of topologically ordered 1d systems, beyond other popular analysis made, for instance, with the string order parameter \cite{string}.
In terms of the properties of the gs manifold, this means that the degeneracy can, yet, be lifted, but violating the symmetry of the system. In this sense the  topological order of the gs is protected by the symmetry \cite{Wen-protected,AKLT-pollmann,AKLT-protected}.
A systematic analysis of spin models with symmetry protected topological order has been performed, and it was demonstrated how the range of entanglement is the crucial property to inspect (long range entanglement seems the inherent property of topologically ordered states in 1d) \cite{Wen-protected}.

In this paper we demonstrate how the 1d cluster states enjoy a symmetry protected topological order as the one we summarized above. Here we mention that such property is originally identified in the Ref.\cite{Son11} and corroborated by \cite{smacchia} as a by product of a problem formulated in cross fertilization area spanned by statistical mechanics, quantum information and cold atoms physics. There the symmetry protection of the cluster states was found by inspection. In the present paper we provide a constructive proof of the symmetry protection.
Customarily, the cluster state is considered as the gs of certain Hamiltonian (the 'cluster hamiltonian') that is the sum of commuting
stabilizer operators in one dimensional lattice \cite{Briegel01}. The gs is unique for periodic boundary conditions, but the state become  fourfold-degenerated if open boundary conditions are applied to the Hamiltonian system; this degeneracy ultimately arises because the completeness of set of stabilizers is spoiled and the free ends spin is not to be stabilized. Then, the system is demonstrated to enjoy a global $Z_2\times Z_2$ symmetry;  different degenerate gs cannot be distinguished by any local operator
commuting with that symmetry. Our approach provides the formalism to obtain the optimal symmetry operators and it has the form of linear sum of two string operators for the 1D cluster state.

\begin{figure}[thbp]
\begin{center}
\includegraphics[width=3in]{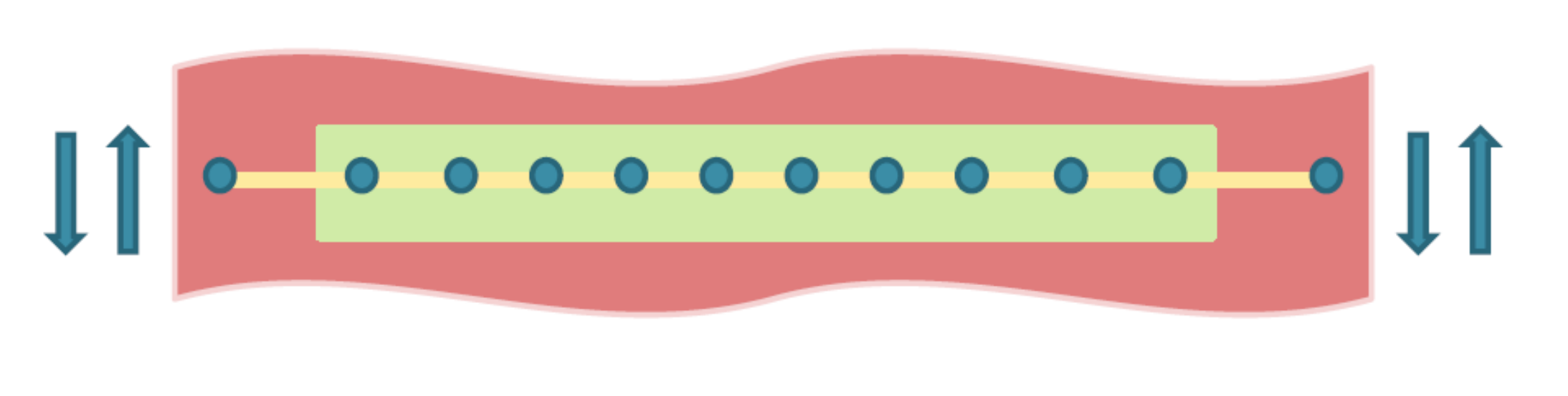}
\caption{\label{fig:1Dspin} One dimensional spin chain of cluster state which is subject to symmetric operation. The state becomes four-fold degenerated due to the edge state and the total state is stabilized by $Z_2\times Z_2$ operation. The symmetric operation can protect gs degeneracy from the local and the quasi-local perturbation.}
\end{center}
\end{figure}


\section{Symmetry protected cluster states}

The 1d cluster states $|C\rangle$ can be constructed as the common eigenstates of a certain set of operators known as stabilzers: $S_i\doteq\sigma_{i-1}^z\sigma_{i}^x\sigma_{i+1}^z $, $S_i |C\rangle=|C\rangle\; , \forall i$ with $\left [ S_i, S_j\right ]=0\; , \forall i,j$ \cite{stabilizers}.
Therefore  $|C\rangle$  is  the gs  of the cluster Hamiltonian
\begin{equation}
	H_C=-\sum_{i=1}^N \sigma_{i-1}^z\sigma_{i}^x\sigma_{i+1}^z \; ,
\label{eq:hamiltonian}
\end{equation}
describing interactions among three spins  located in next nearest neighbors  of the one dimensional lattice $\Lambda$; $\sigma_i^\alpha$ are Pauli matrices. For periodic boundary conditions the system (\ref{eq:hamiltonian}) is integrable by construction because the set of $S_i$ define a complete set of commuting operators. In such a case the gs is unique.
For open boundary conditions $\sigma_0^z =\sigma_{N+1}^z=0$, instead,
the gs is $4$-fold degenerate because two stabilizers are removed from the complete set:
\begin{equation}
\label{eq:cluster}
	\mathcal C_0 =\mbox{span} \left\{ \frac{1}{2^N}(\sigma^z_1 )^k(\sigma^z_L)^l \prod_{i} \hat{C}_i
	|+\rangle \; \; k,l=0,1\right\}
	\label{eq:C0}
\end{equation}
where ${C}_i=\left (\id
	+\sigma_i^z+\sigma_{i+1}^z-\sigma_i^z \sigma_{i+1}^z \right )$ and $\sigma^x|+\rangle = |+\rangle$.
	 The four states obtained by $\{k,l\}=\{0,1\}$ span the gs manifold (the case of periodic boundary conditions, $\sigma_0^z =\sigma_{N}^z$
and $\sigma_{N+1}^z=\sigma_1^z$,corresponds to  $k=l=0$). We observe that vectors in such a manifold can be distinguished by a certain set of local operators $O_{loc}$. This would spoil any topolgical feature out of cluster states because we could perturb  $H_C$ with $O_{loc}$ to lift the degeneracy of the ground state.
In the present case the operators $O_{loc}$ are elements of the set $\Sigma$ obtained combining the generators of the compact group
\begin{equation}
\label{eq:generator}
G=\{\sigma^z_1, \sigma^z_L, \sigma^x_1\sigma^z_2, \sigma^z_{L-1}\sigma^x_L\}
\end{equation}
in an arbitrary way (i.e. perturbing the Hamiltonian $H$ by $\Sigma$ stabilizes the gs). This ultimately arise because  $H_C$ is constructed as a linear sum of stabilizers. By the application of the control phase gates $C_i$, the stabilizers, in turn, can be reduced into the simple Pauli operators spanning the qubit's Hilbert space (namely the  representation space of  $s=1/2$ $su(2)$ algebra).

It is instructive to look at the problem in the space of the Pauli algebra $\{\tau^\alpha\}$  obtained  after the control-phase unitary transformation is applied to the $\sigma^\alpha$:
\begin{eqnarray}
&&U_{cp}  \sigma^x_i U_{cp} =\tau_{i-1}^z\tau_{i}^x\tau_{i+1}^z\;,\,  U_{cp} \sigma^y_i U_{cp}  =\tau_{i-1}^z\tau_{i}^y\tau_{i+1}^z\;, \nonumber \\
&&U_{cp}\sigma^z_i U_{cp} =\tau_{i}^z\; ,
\end{eqnarray}
where $U_{cp}\doteq \left (\prod_i C_i\right )$.
In this basis $G\rightarrow \bar{G}$ where $\bar{G}=\{\tau^z_1, \tau^z_L, \tau^x_1, \tau^x_L,\}$ and $\bar{G}$ is the compact set of the group generators in the spin space at sites $1$ and $L$. Therefore, only the group elements generated by $\bar{G}$ are stabilizing the gs.
We comment that many local operators can be a relevant perturbation of the  energy spectrum of  $H_C$ (like  $\{\sigma_i^x, \sigma_i^x\sigma_j^x, \sigma_i^z, \sigma_i^z\sigma_j^z, \sigma_i^x\sigma_j^z, 2\leq i, j\leq L-1 \}$)  but it can be proved that they cannot affect the degeneracy of gs. Despite of the fact of spectrum changes, we will also investigate the symmetry which can protect from all the quasi-local perturbation.

It is worth noting that the parity symmetry $\prod_i \sigma^z_i$ and $\sigma^y_1\prod_{j=2}^{L-1}\sigma^z_j \sigma^y_L $ are the set which do not commute with all the generators of $G$ and therefore they protect the gs degeneracy by perturbing $H_C$ by a linear combination of them. However the products of generators of $G$ indeed commute with the symmetries above, and therefore operators like $\sigma^z_1\sigma^z_L$,  $\sigma^x_1\sigma^z_2 \sigma^z_{L-1}\sigma^x_L$ etc. can be taken as 'legitimate' terms destroying the gs degeneracy. We will now demonstrate how the dangerous operators
can be excluded by constructing the symmetries of $H_C$ as linear combinations of operators acting on $\otimes_{i=1}^L {\cal H}_i $ Hilbert space.
Our aim is to promote the two stabilizers $S_1$ and $S_2$ to {\it global} symmetries $T_1$ and $T_2$, $\left [H_C, T_i\right ]$, protecting the
gs degeneracy: $\left [ T_i,\Sigma\right ]\neq 0$.
We will obtain that $|T_i|^2=1$, demonstrating that the system (\ref{eq:hamiltonian}) enjoys indeed a ${\cal Z}_2 \times {\cal Z}_2$ symmetry.


To illustrate the underlying philosophy leading to the construction of $T_1$ and $T_2$, we first refer to operators spanning a local symmetry. For the symmetry, conditions of a local operator $T^{loc}$ are to satisfy $[T^{loc}, \sigma_1^{\alpha}]\neq 0$,
$[T^{loc}, \sigma_L^{\alpha}]\neq 0$ and  $[T^{loc}, \sigma_1^{\alpha}\sigma_L^{\alpha}]\neq 0$ where $\alpha \in\{x,y,z\}$. The relations can be fulfilled simultaneously only if $T^{loc}$ is a linear combination of the elements of Pauli group.
%
In the present case it suffices to consider $T^{loc}_s= \frac{1}{\sqrt{2}}\left(A^{loc}_s +B^{loc}_s\right) \;, s=1,2$, $A^{loc}_s$ and $B^{loc}_s$ being local elements of the Pauli group; additionally we require: (i)
$|T^{loc}_s|^2=1$ implying that $\{A^{loc}_s, B^{loc}_s\}=0$.
(ii)   $[A^{loc}_1, A^{loc}_2] = [B^{loc}_1, B^{loc}_2]=0$,
$[A^{loc}_1, B^{loc}_2] = [A^{loc}_1, B^{loc}_2]=0$, which guarantees that $[T^{loc}_1, T^{loc}_2]=0$.
The explicit form of  $A_s^{loc}$ and  $B_s^{loc}$ can be fixed requiring that such operators do not commute with any element of $\Sigma$ simultaneously (for the same $s$).
Here, it is convenient to exploit the controlled phase basis for which $G\rightarrow \bar{G}$.
We observe that the two operators  $(\tau_1^x\tau_L^y, \tau_1^z\tau_L^y)$ are not simultaneously commuting with any operator element in $\bar{G}$ neither with the combinations of the elements in $ \bar{G}$: $\tau_1^\alpha\tau_L^{\beta}$ where $\alpha, \beta\in\{z,x\}$.
Therefore a  local ${\cal Z}_2$ can be spanned by  $(\bar{A}_1^{loc}, \bar{B}_1^{loc})=(\tau_1^z\tau_L^y, \tau_1^x\tau_L^y)$.
To generate the  second local copy of ${\cal Z}_2$ (commuting with the above ${\cal Z}_2$)  we consider  the operators $(\tau_L^y, \tau_1^y\tau_L^z)$ \cite{note}.
Transforming back into the operational basis, the operators $A^{loc}_s$ and $B^{loc}_s$ for $T^{loc}_s$ are finally given by
$
\left(A^{loc}_1,  B^{loc}_1\right)=\left(\sigma_1^x\sigma_2^z\sigma_{L-1}^z\sigma_L^y,~\sigma_1^z\sigma_{L-1}^z\sigma_L^y\right) $, $\left(A^{loc}_2, B^{loc}_2\right)=\left(\sigma_{L-1}^z\sigma_L^y,~\sigma_1^y\sigma_{2}^z\sigma_L^z \right).
$ 


Below we construct the symmetries  $T_1$ and $T_2$ promoting the procedure depicted above from a local to  a global level.
To stabilize the state to any local perturbation through the lattice, the operators $T_1$ and $T_2$ are required to
satisfy: $[T_i, \sigma^\alpha_j]\neq 0,~ \forall j,~ 2\leq j\leq L-1$, $\alpha\in \{x, y, z\}$.
In that case, the number of qubits for the symmetry is constrained by multiples of odd numbers as it will be cleared in the following discussions.
In the current case of open cluster state, the global operators $(A_1, B_1)$ and $(A_2, B_2)$ read as
\begin{eqnarray}
\nonumber
A_1 &=&\prod_{n=0}^{\frac{L-9}{6}} \Big( \sigma_{6n+1} ^y \sigma_{6n+2} ^x  \sigma_{6n+3}^x \sigma^y_{6n+4}  \sigma_{6n+5} ^z \sigma_{6n+6} ^z \Big) \sigma_{L-2}^y\sigma_{L-1}^x\sigma_{L}^x \\
B_1&=&\prod_{n=0}^{\frac{L-9}{6}} \Big( \sigma_{6n+1} ^z \sigma_{6n+2} ^z  \sigma_{6n+3}^y \sigma^x_{6n+4}  \sigma_{6n+5} ^x \sigma_{6n+6} ^y \Big) \sigma_{L-2}^z\sigma_{L-1}^z\sigma_{L}^y \\
\nonumber
A_2&=&\prod_{n=0}^{\frac{L-9}{6}} \Big( \sigma_{6n+1} ^z \sigma_{6n+2} ^y  \sigma_{6n+3}^x \sigma^x_{6n+4}
\sigma_{6n+5} ^y \sigma_{6n+6} ^z \Big) \sigma_{L-2}^z\sigma_{L-1}^y\sigma_{L}^x\\
B_2&=&\prod_{n=0}^{\frac{L-9}{6}} \Big( \sigma_{6n+1} ^y \sigma_{6n+2} ^z  \sigma_{6n+3}^x \sigma^y_{6n+4}  \sigma_{6n+5} ^x \sigma_{6n+6} ^x \Big) \sigma_{L-2}^y\sigma_{L-1}^z\sigma_{L}^z
\end{eqnarray}
These operators obviously square to the identity: $A_1^2=A_2^2=B_1^2=B_2^2=I$ and,
if $L = 3(2k+1)$, with $k$ positive nonzero integer, the following anticommutation relations hold:
$\{A_1,B_1\}=\{A_2,B_2\}=0$ and commute otherwise. The length $L$ of the string operators $A_i$ and $B_i$ are the integer multiples of 6 plus 3 whose structure will become obvious shortly. In any case, they span the carrier
space for the first and $L$-th copy of $su(2)$ algebra representation (thus providing
a realization of the logical operators in the ground space manifold).
Therefore
\begin{equation}
T_1=A_1+B_1\quad , \quad T_2=A_2+B_2
\end{equation}
generate the desired $Z_2\times Z_2$ indeed protecting the degeneracy of the gs and therefore the operator constitute all the conditions for the topological order in the gs of $H_C$.

Additional insights are achieved in the control phase basis, namely by the application of the control phase gate operation $U_{cp}$: $\bar{T}_s \doteq U_{cp} T_s U_{cp}$. 
In the new basis, the transformed symmetry operators read
\begin{eqnarray}
\label{symmetry_cp}
\bar{T}_1=\bar{A}_1 +\bar{B}_1 &&=
\tau^x\tau^x\tau^x\tau^x\id \id~ \tau^x\tau^x\tau^x\tau^x  \id \id~ \cdots  \tau^x\tau^x \tau^y \nonumber \\
\hspace*{-3cm} &&+\tau^z \id \tau^x\tau^x\tau^x\tau^x  ~\id \id\tau^x\tau^x\tau^x\tau^x ~ \cdots \id \id \tau^y
\nonumber\\
 \bar{T}_2=\bar{A}_2 +\bar{B}_2 &&=
\id \tau^x\tau^x\tau^x\tau^x  \id~ \id \tau^x\tau^x\tau^x\tau^x  \id ~ \cdots \id \tau^x \tau^y \nonumber \\
&&+ \tau^y \id \id \tau^x\tau^x\tau^x~ \tau^x  \id \id \tau^x\tau^x\tau^x~ \cdots \tau^x\id \tau^z \nonumber  \\
\end{eqnarray}
One can observe that a sequence of six operators, four $\tau^x$ followed by two $\id$, are periodically appeared in the middle of local edge operators.
We comment that the specific sequence of the six operators in (\ref{symmetry_cp}) results from control phase operations on the $\sigma^\alpha$
producing stabilizer operators in the $\tau$-basis. Because of such sequence $\bar{A}_s$ and the $\bar{B}_s$ cannot commute with the operators
$\tau^\alpha$, simultaneously. This implies, in turn, that the operators $\bar{T}_1$ and $\bar{T}_2$
cannot commute with any of the operators generated by the operator set $\bar{G}$.
In the $\sigma$-basis this  leads to  symmetry protection against any local perturbation: $[T_i, \sigma^\alpha_j]\neq 0,~ \forall j,~ 2\leq j\leq L-1$, $\alpha\in \{x, y, z\}$.

\section{Perturbation of the cluster states } An important question to ask is whether a topological phase  can originate from the cluster ground state by perturbation.
For generic perturbation the degeneracy of the gs is lifted. Topological order in one dimensional systems, however, is not necessarily implied by gs degeneracy. An emblematic example is provided by the Haldane phase resulting from the AKLT gs perturbed by a magnetic field\cite{haldane,gu-wen}. In that context it was proved that the stability of the hidden order in the Haldane phase arises because the symmetries of the system forbid to deform it to a fully polarized phase. For the cluster phase a similar scenario emerges. Indeed we observe that the Hamiltonian in (\ref{eq:hamiltonian}) (for periodic boundary conditions) is translationally invariant, and enjoys a $Z_2\times Z_2$ symmetry spanned by the parity symmetry $\sigma^z\rightarrow \sigma^z$,
$\tau^{x,y}\rightarrow -\sigma^{x,y}$ and by the time reversal symmetry  $\sigma^{x,z}\rightarrow \sigma^{x,z}$,
$\sigma^{y}\rightarrow -\sigma^{y}$. We observe that the projective representation of $Z_2\times Z_2$ is the Pauli algebra. Relying on \cite{gu-wen}, this suggest that any perturbation fulfilling that symmetry lead the system to a phase with non-trivial topological order (distinct from the fully polarized phase). This is the case considered
in Refs.\cite{Son11,smacchia} (see also \cite{Bartlett} for similar models) where the cluster Hamiltonian was perturbed by $H_I=\lambda \sum_i \sigma^y_i \sigma^y_{i+1}$. There it resulted that an extended gapped
phase originates from the cluster ground state that is characterized by a non local correlation marked by a non vanishing string order. Here we provide further evidence that
such a phase is provided by a topological order that protected by parity and time reversal symmetries. The model $H=H_C+H_I$ display a quantum phase transition at $\lambda=1$ where  the parity symmetry is broken; the corresponding gapped phase does not display any topological feature because the projective representation of the remaining $Z_2$ is trivial.

\section{Conclusion}
We provide the explicit construction of the $Z_2\times Z_2$ symmetry which protects the topological order of the cluster state under local perturbation. The symmetry results  particularly transparent in the transformed basis given by control phase operations. We found that the topological order in the cluster ground state is robust against the perturbation with an Ising interaction since it is protected by parity $\&$ time reversal symmetry. Our formalism is general and, in principle,  can be applied to different spin models. However, the implication of our approach to 2D cluster states remains to be investigated further. The formalism developed in \cite{Nussinov,Cobanera} might provide  a useful platform to study the problem.  We finally comment that symmetry protection might be a route to preserve topological order by thermal fluctuations\cite{Nussinov}.


\bigskip 

\begin{acknowledgements}
We thank A. Hamma for discussions.
\end{acknowledgements}




\begin{thebibliography}{99}
\bibitem{note}
 It is worth noting  that, for given  $(\bar{A}_1^{loc}, \bar{B}_1^{loc})$ such a choice  is not unique. However the other possible legitimate choice $(\tau_1^y\tau_L^z, \tau_1^y\tau_L^x)$, $(\tau_L^y, \tau_1^y\tau_L^x)$,  $\tau_L^y$, $\tau_1^y\tau_L^z$ and $\tau_1^y\tau_L^x$ at a local level would lead to global symmetries that cannot protect the gs by local perturbations.
\end{thebibliography}
\end{document}